\documentclass[12pt]{article}
\usepackage{a4}
\usepackage{axodraw}
\usepackage{cite}
\usepackage{graphicx}
\usepackage{amssymb}
\usepackage{amsmath}
\usepackage{epsfig}
\usepackage[tmargin=1cm,bmargin=1cm,lmargin=2cm,rmargin=2cm]{geometry}

\newcommand{\qb}{\ensuremath{\overline q}}

\newcommand{\be}{\begin{equation}}
\newcommand{\bdm}{\begin{displaymath}}
\newcommand{\bea}{\begin{eqnarray}}
\newcommand{\beastar}{\begin{eqnarray*}}

\newcommand{\ds}{\ensuremath{\delta_s}}
\newcommand{\dc}{\ensuremath{\delta_c}}

\newcommand{\ee}{\end{equation}}

\newcommand{\e}{\ensuremath{\epsilon}}
\newcommand{\eps}{\ensuremath{\epsilon}}

\newcommand{\edm}{\end{displaymath}}
\newcommand{\eea}{\end{eqnarray}}
\newcommand{\eeastar}{\end{eqnarray*}}

\newcommand{\mur}{\ensuremath{\mu_{R}}}

\newcommand{\Msq}{\ensuremath{ { \overline {|M^{(0)}|^2} }}}

\newcommand{\rarrow}{\ensuremath{\rightarrow}}
\newcommand{\ReDs}{\ensuremath{{\cal R}e {\cal D}_s}}
\newcommand{\ImDs}{\ensuremath{{\cal I}m {\cal D}_s}}

\newcommand{\s}{\ensuremath{\sigma}}

\newcommand{\cd}{\ensuremath{\mathcal D}}

\newcommand{\ct}{\ensuremath{\mathcal T}}

\nopagebreak

\begin{document}
\begin{flushright}
TIFR/TH/10-11   \\ 
\end{flushright}
\baselineskip 22pt
\begin{flushright}
\end{flushright}
\vskip 45pt
\begin{center}
{\large \bf
$W^+W^-$ production in large extra dimension model at next-to-leading order
in QCD at the LHC
} 
\\
\vspace{8mm}
{\bf
Neelima Agarwal$^a$
\footnote{neel1dph@gmail.com},
}
{\bf
V. Ravindran$^b$
\footnote{ravindra@hri.res.in},
\\
Vivek Kumar Tiwari$^a$
\footnote{vivekkrt@gmail.com},
Anurag Tripathi$^c$
\footnote{anurag@theory.tifr.res.in}
}\\
\end{center}
\vspace{10pt}
\begin{flushleft}
{\it
a)~~Department of Physics, University of Allahabad, Allahabad 211002, India. \\
b)~~Regional Centre for Accelerator-based Particle Physics,\\
~~~~Harish-Chandra Research Institute, Allahabad 211019, India.\\
c)~~Department of Theoretical Physics, Tata Institute of Fundamental Research, \\ 
~~~~Mumbai 400005, India. \\
}
\end{flushleft}
\vspace{10pt}
\centerline{\bf Abstract}
We present next-to-leading order QCD corrections to production of two $W$ bosons  
in hadronic collisions in the extra dimension ADD model.
Invariant mass and rapidity distributions are obtained to order $\alpha_s$ in QCD by 
taking into account all the parton level subprocesses. The computation is 
organized using the monte carlo  based method of phase space slicing.
We estimate the impact of the QCD corrections on various observables and find that they
are significant.  
We present some results for a $10~TeV$ LHC
but most of the results presented here are for $14~TeV$ LHC.
We also show the reduction in factorization scale uncertainty when
${\cal O}(\alpha_s)$ effects are included. 

\vskip12pt
\vfill
\clearpage

\setcounter{page}{1}
\pagestyle{plain}

\section{Introduction}
\newcommand{\vn}{\ensuremath{{\vec n}}}
The fact that electroweak symmetry breaking scale of the Standard Model (SM) 
cannot be made stable against the quantum corrections (hierarchy problem) 
within the SM indicates to the possibility of new physics at TeV scale.
The Large Hadron Collider (LHC) which will operate at an enormous
center of mass energy (${\sqrt S}=14 TeV$) offers to shed light on
the existence of new physics. The most popular new physics models are
based on the ideas of supersymmetry and extra spatial dimensions.
Proposals to address the hierarchy problem using 
extra dimensions were introduced in 
\cite{Antoniadis:1998ig,Randall:1999ee}.
In this paper we will consider the model by 
Arkani-Hamed, Dimopoulos and Dvali (ADD)
\cite{Antoniadis:1998ig}. 
In this model all SM fields are confined to a (3+1) dimensional manifold 
and the extra $d$ spatial dimensions are 
compactified, with same radius of compactification $R$, on a $d$-torii.  
The effect of extra dimensions appears as Kaluza Klein (KK) gravitons
on the 3-brane which couple to SM fields through energy momentum tensor
with a strength $\kappa$ which is 
related to the volume of the extra dimensions, 
and the fundamental scale $M_s$ in $4+d$ dimensions by \cite{Han:1998sg}
\begin{equation}
\kappa^2 R^d= 8\pi (4\pi)^{d/2}\Gamma(d/2)M_s^{-(d+2)}.
\label{mass}
\end{equation}
Although the coupling $\kappa$ is $M_{Pl}$ suppressed, the fact that there are 
large number of KK modes that couple to the SM fields makes the 
cumulative effect significant and leads to observable effects.  
One extra dimension ie. $d=1$ is ruled out
\cite{Kapner:2006si}
and $d=2$ is severely constrained  so we will consider
in this paper $d=3$ and above.
There are two ways to probe such effects at colliders, either through gravitons emission 
or by virtual graviton exchange.
In this paper we will consider only the effects of virtual spin-2 KK states.

The precise measurement of hadronic production of gauge boson pairs 
is one of the important endeavors at the LHC both in the context of SM and new physics
studies. Studies in other channels have been reported in  
\cite{Mathews:1999iw}
in extra  dimension models.
In this paper we will consider production of $W$ pair at the LHC. 
Owing to its importance, its study 
has attracted a lot of attention in the literature. Many studies have been 
carried out for its production in the SM; a study in the context  
of anomalous triple gauge boson vertices was carried out in  \cite{Hagiwara:1986vm}.
Leading order (LO) studies in the SM can be found
in \cite{Brown:1978mq}.  
As is well known the LO results are highly sensitive to the arbitrary 
renormalization and factorization scales. At this order the 
factorization scale $\mu_F$ enters solely through the parton distribution
functions as the parton level cross-section, at this order,
does not depend on $\mu_F$. 
As we include higher order terms of the perturbation series the dependence
will reduce and an all order result will be completely independent of these 
arbitrary scales. In addition the NLO results are usually 
significantly enhanced as compared to the LO results. It is thus important to carry out NLO calculation to 
reduce sensitivity to these scales.
Because of its importance, $W^+W^-$ production has been studied to 
next-to-leading-order (NLO) accuracy
in the SM \cite{Ohnemus:1991kk,Ohnemus:1994ff}.
It has also been studied via gluon fusion through a quark box loop
or triangle quark loop with $\gamma$ or $Z$ boson exchange
\cite{Kao:1990tt}
and at one and two loop levels in high energy limit in SM 
\cite{Chachamis:2008xu}.
The significance of NLO computations in the extra dimension models for  
Drell-Yan \cite{Mathews:2004xp},
diphoton \cite{Kumar:2008pk},
ZZ \cite{Agarwal:2009xr}, 
graviton+photon \cite{Gao:2009pn}, graviton+jet \cite{Karg:2009xk}
production has already been demonstrated.
These studies show that not only the predictions at NLO are enhanced
but are also less sensitive to the factorization scale.
With this in mind we carry out a 
complete NLO calculation of hadronic $W^+W^-$
production in ADD model in this paper. $W^+W^-$ production in Randall Sundrum model
is presented in our work \cite{Agarwal:2010sn}.

We organize the paper in the following sections as follows.  
In section $2$ we give the details of NLO computation for the $W^+W^-$ production. Here
we give the matrix element squares for all the $2 \rightarrow 2$ subprocesses
both at leading order and at loop level at order $a_s~ (=g_s^2/16 \pi^2)$. All 
the Feynman diagrams are collected in the appendix. 
The virtual corrections  contain soft and collinear singularities
which appear as simple and double poles in $\epsilon$ as we use
dimensional regularization with $n=4+\epsilon$.
The $2 \rightarrow 3$ real emission
matrix elements which were also calculated analytically are not 
presented here (to save space as the expressions are 
lengthy) and can be obtained
on request. These computations were done using
the symbolic manipulation program FORM \cite{Vermaseren:2000nd}.
We intend to use monte carlo methods for obtaining kinematical 
distributions. These methods prove very useful if experimental
cuts need to be imposed on the final state particles and they 
avoid the need for repeating calculations for obtaining
different distributions as would be required by completely 
analytical methods. In this paper we employ the method of 
two cutoff phase space slicing \cite{Harris:2001sx} to carry out
our calculation. This method gives numerically stable results
as has been demonstrated in \cite{Harris:2001sx} and also
in our previous works 
\cite{Kumar:2008pk, Agarwal:2009xr}.
An alternative to this is dipole subtraction 
method \cite{CataniSeymour} which is also widely used in higher order QCD
calculations. We describe, in brief, phase space slicing method and 
how the soft
and collinear singularities that appear at virtual level and
real emission level are treated. In section 3 we present some checks
on our code and then present some useful kinematical distributions.

\section{Next-to-leading-order Computation}

The hadronic production of $W$ bosons at NLO has three pieces of computation.
A LO piece which is a $2 \rarrow 2$ parton level process  
($q \bar{q} \rightarrow W^+W^- $ and $ gg \rightarrow W^+W^- $);  
second is the $2 \rarrow 2$ order $a_s $ 
piece which originates from loop corrections to the LO piece; the third and final part 
originates from real emission processes where in addition to two $W$ bosons, a parton is
also emitted in the final state 
($q \bar{q} \rightarrow W^+W^-g,~  
qg \rightarrow W^+W^-q $,
and $ gg \rightarrow W^+W^-g $ ). 
Let us take up these three pieces in turn.

\subsection{Born Contribution}

\begin{figure}[ht]
\centerline{\epsfig{file=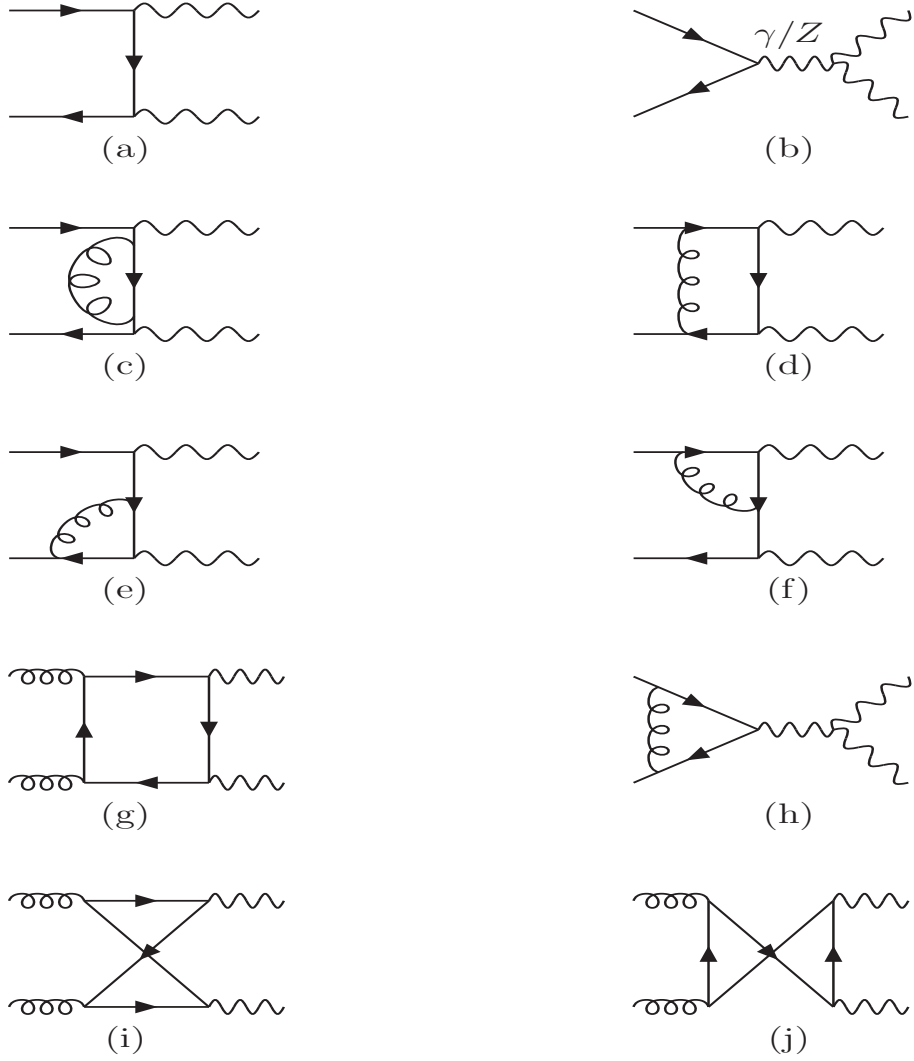,width=12cm,height=14cm,angle=0} }
\caption{Leading order and order $a_s$ virtual diagrams in SM for the subprocess
  $ u\bar{u}\rightarrow W^+W^-$ and $ gg \rightarrow W^+W^-$. 
  The diagrams for the subprocess 
  $ d\bar{d}\rightarrow W^+W^-$ are obtained by replacing $u \rightarrow d$
  and $W^+ \leftrightarrow W^-$ in the diagrams shown here.}
\label{smvrt}
\end{figure}

The charged vector-boson production in the leading partonic scattering processes corresponds to
\be
a(p_1) + b(p_2) \rarrow W^+(p_3) + W^-(p_4).
\ee
where $p_1$ and $p_2$ are the momenta of initial state partons while $p_3$ and $p_4$ are 
those of final state vector bosons.
The $W$ boson pair can couple to $KK$ gravitons, so it is possible 
to produce them through virtual graviton exchange 
at the leading order \cite{Agashe:2007zd}. 
In SM this proceeds via t channel (or u channel) quark anti-quark annihilation
along with s-channel $\gamma$ and $Z$ boson exchange shown in Fig.~\ref{smvrt}. 
The coupling of fermions to $W$ and $Z$  bosons are respectively
\be
-i \frac{e \ct_W}{2} \gamma^{\mu} \left(1 - \gamma^5 \right) ,  \quad \quad
-i \frac{e {\cal T}_Z}{2} \gamma^{\mu} \left(C_v - C_a\gamma^5 \right).
\ee
where $\cal T_W$ and $\cal T_Z$ read 
\be
{\cal T}_W = \frac{1}{\sqrt{2} \sin\theta_W } ~, 
\quad \quad 
{\cal T}_Z= \frac{1}{\sin\theta_W \cos\theta_W}.
\ee
The coefficients $C_v$ and $C_a$ are given by 
\bea 
 C_v = T_3^f - 2~ Q_f \sin^2\theta_W,  \quad \quad C_a = T_3^f,  \quad \quad
\eea
The $Q_f$ and $T_3^f$ denote the electric charge and the third component of 
the weak isospin of the fermion $f$ respectively, $\theta_W$ is the weak mixing angle and
$m_z$ is the mass of $Z$ boson.
The $Z$ boson propagator is given as
\be
\frac{-i g_{\mu\nu}}{s-m_z^2 +i ~\Gamma_z m_z} .
\ee
We have used unitary gauge in the electroweak sector ie. $\xi=\infty$, this
simplifies the calculation as both the goldstone boson and ghosts in 
the electroweak sector disappear.

The $\gamma_5$ matrices that appear in the intermediate stages of the 
computation require special care as they are not defined in arbitrary dimensions. 
We have used naive anti-commutation relations between
$\gamma_5$ and other gamma matrices in $n$ dimensions and the resulting
traces are then computed in $n$ dimensions as they are free of $\gamma_5$.
Alternatively, one can use other method
namely HVBM-scheme which was proposed in \cite{'tHooft:1972fi} and generalized in \cite{BM}.
In this approach, Gamma matrices and momenta in the loop and final state phase space integrals
are split into a 4 and an $n-4$ dimensional part.  The $\gamma_5$ anti-commutes in 4 dimensions
and commutes in $n-4$ dimensions with rest of the $\gamma$ matrices.

\begin{figure}[ht]
\centerline{\epsfig{file=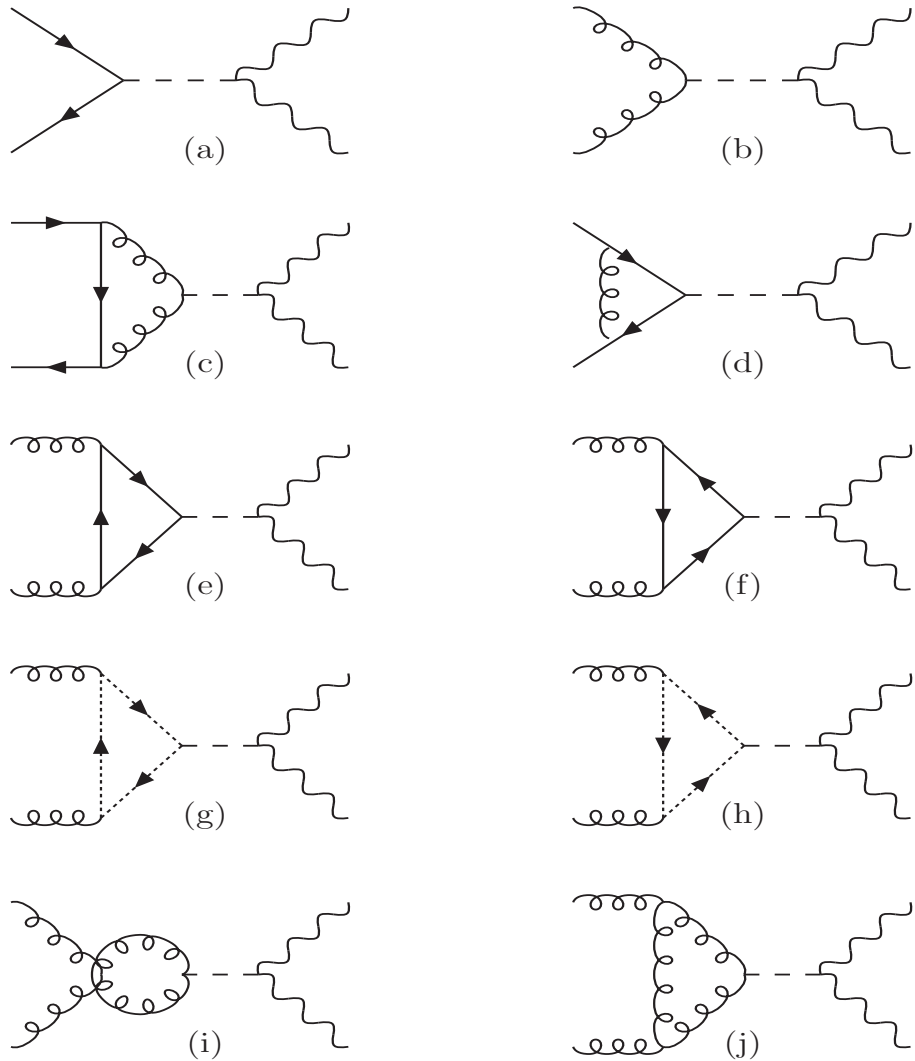,width=12cm,height=14cm,angle=0} }
\caption{Leading order and order $a_s$ gravity mediated virtual correctionss for the subprocess
  $ q\bar{q}\rightarrow W^+W^-$ and $gg \rightarrow W^+W^-$. 
The big dashed
lines represent gravitons and the small dashed lines represent QCD ghosts.} 
\label{bsmvrt}
\end{figure}

We give below the matrix element squares summed (averaged)
over the final (initial) state spins, colors and polarizations notated 
by an overline
in $n=4+\eps$ dimensions . s, t and u are the usual Mandelstam invariants. 
We will denote by $sm$, $gr$ and $int$  the contributions from SM,
gravity, and interference of SM with gravity respectively. 
The SM at  LO gives order $e^4$ contribution to the cross sections:
\bea
\label{Msmlo}
\Msq_{q \qb, sm} &=& \frac{e^4}{N} \left(A_1^q B_1^q +A_2^q B_2^q +A_3^q B_3^q\right)  
\eea
where
\bea
\label{coup}
A_1^u &=& {\cal T}_W^4 ~,                                                    
\nonumber \\ [2ex]
A_2^u &=& \frac{e_q^2}{s^2}
+ e_q~ C_v~ {\cal T}_Z~ \cot\theta_W 
\frac{ (s-m_z^2)}{s \left[(s-m_z^2)^2+\Gamma_z^2 m_z^2\right]}           
\nonumber \\ [2ex]
&&
+ (C_v^2+C_a^2)~{\cal T}_Z^2~ \cot^2\theta_W
 \frac{1} {4 \left[(s-m_z^2)^2+\Gamma_z^2 m_z^2\right]}  ~, 
\nonumber \\ [2ex]
A_3^u &=& \frac{e_q {\cal T}_W^2}{s} 
+ (C_a+C_v)~ {\cal T}_W^2~ {\cal T}_Z~ \cot\theta_W 
\frac{(s-m_z^2)}{2 \left[(s-m_z^2)^2+\Gamma_z^2m_z^2\right]}~. 
\nonumber \\ 
\eea
$A_i^q$ are in their essence combinations of  EW couplings and 
propagator factors and $B_i^q$ are the functions of the 
kinematic invariants. $N$ denotes the number of colors. 
In case of up type quark initiated processes,
$B_1^u$ originates from purely
t-channel, $B_2^u$ from purely s-channel  while $B_3^u$ from 
the interference of t and s-channel diagrams. These functions are
given by
\bea
\label{funcb}
B_1^u &=&   \frac{1}{4 m^4 t^2}     
           \Bigg[-\left\{m^8\left(-2 + n\right)^2\right\} + t^3u - 
            2m^2\left(-2 + n\right)t^2\left(t + u\right)  
\nonumber \\ [2ex]
&&
           + m^4t\left\{\left(-9 + 4n\right)t + \left(-2 + n\right)^2u\right\}\Bigg] ~, 
\eea
\bea 
B_2^u &=&       \frac{1}{2 m^4}
            \Bigg[-20m^8\left(-2 + n\right) + 8m^6\left(-2 + n\right)\left(t + u\right) + 
            tu\left(t + u\right)^2 
\nonumber \\ [2ex]
&&
             - 2m^2\left(-2 + n\right)\left(t + u\right)^3 
             + m^4\Big\{\left(-9 + 4n\right)t^2 + 2\left(-13 + 6n\right)tu 
\nonumber \\ [2ex]
&&
             + \left(-9 + 4n\right)u^2\Big\}\Bigg] ~,
\eea
\bea 
B_3^u &=&  \frac{1}{2 m^4 t}
            \Bigg [-10m^8\left(-2 + n\right) + 4m^6\left(-2 + n\right)\left(t + u\right) + 
            t^2u\left(t + u\right) 
\nonumber \\ [2ex]
&&
             - 2m^2\left(-2 + n\right)t\left(t + u\right)^2 + 
            m^4t\left\{\left(-9 + 4n\right)t + \left(-13 + 6n\right)u\right\}\Bigg] ~. 
\eea
Here $m$ denotes the mass of final state $W$ bosons.
The $B_i^q$ expressions for down type quarks are related to that of up type quarks
as follows
\bea
B_1 ^d \left(t,u,s \right) &=&  B_1^u \left(u,t,s \right) ~, 
\nonumber \\ [2ex]
B_2 ^d \left(t,u,s \right) &=&  B_2^u \left(t,u,s \right) ~,
\\ [2ex]
B_3 ^d \left(t,u,s \right) &=& -B_3^u \left(u,t,s \right) ~.
\nonumber
\eea
In addition, two more processes are allowed as the $KK$ gravitons can appear at the 
propagator level, $q\qb \rarrow G^* \rarrow WW$ and $gg \rarrow G^* \rarrow WW$, as 
shown in Fig.~\ref{bsmvrt}. As we use unitary gauge in the electroweak sector
the term proportional to $1/\xi$ in the $WW-$graviton vertex 
\cite{Han:1998sg} drops out. 
The $q \qb$ and $gg$ initiated contributions which are of order $\kappa^4$ are given below. 
\bea
\label{Mbsmqlo}
\Msq_{q\qb, gr}  &=&
 \frac{1}{64 N} |{\cal D}_s|^2 \kappa^4 \Bigg[
         n \Big\{8 m^8 - 16 m^6 (t + u) + t u (3 t^2 + 2 t u + 3 u^2)
\nonumber \\[2ex]
&&            + m^4 (9 t^2 + 30 t u + 9 u^2)
            - 2 m^2 (t^3 + 7 t^2 u + 7 t u^2 + u^3)
           \Big\}
\nonumber \\[2ex]
&&
 - \Big\{8 m^8 - 24 m^6 (t + u) + t u (7 t^2 + 10 t u + 7 u^2)
\nonumber \\[2ex]
&&           + m^4 (17 t^2 + 62 t u + 17 u^2)
           - 4 m^2 (t^3 + 9 t^2 u + 9 t u^2 + u^3) \Big\}
        \Bigg] ~,
\eea
\bea
\label{Mbsmglo}
\Msq_{gg,gr} &=&
\frac{|{\cal D}_s|^2 \kappa^4}{(N^2-1)}
  \frac{1}{128} \times
 \Bigg[ 128 m^8 +9t^4 + 28 t^3 u + 54t^2 u^2 +28 tu^3 + 9u^4
\nonumber \\[2ex]
&&
       -256m^6(t+u)
       +192m^4 (t+u)^2 -64 m^2 (t+u)^3
     - \frac{72}{\left( n-1\right)^2} s^3 (4m^2-s)
\nonumber \\[2ex]
&&
     - \frac{3}{ n-1} s^2 \Big\{ 188m^4 -17t^2 -226 tu -17u^2 +60m^2(t+u) \Big\}
\nonumber \\[2ex]
&&
     + \frac{32}{(n-2)^2} \Big\{ -44m^8 +40m^6 (t+u) -40 m^2 tu(t+u) +9tu(t+u)^2
\nonumber \\[2ex]
&&
                                 + m^4(-9t^2+26tu -9u^2)
                           \Big\}
     + \frac{4}{n-2}
       \Big\{ 692 m^8 -13t^4 -196t^3u -362t^2u^2
\nonumber \\[2ex]
&&
              -196tu^3 -13u^4
              -544 m^6(t+u) -8m^4 (t^2+83tu+u^2)
\nonumber \\[2ex]
&&
            +16m^2(5t^3 +53t^2u +53tu^2 +5u^3)
       \Big\}
 \Bigg] ~.
\eea
We have denoted the sum of spin-2 KK graviton propagators
by $\cd_s$, then $\cd_s$ times square of the coupling can be written as
\be
\label {prop}
\kappa^2 \cd_s =  \frac{8 \pi}{i M_s^4} \left( \frac{\sqrt s}{M_s} \right)^{d-2}
                              \left[ \pi + 2 i I(\Lambda/{\sqrt s}) \right]
\ee
The function $I(\Lambda /\sqrt s)$ depends on the ultraviolet cutoff $\Lambda$ on the
KK modes and its expression can be found in
\cite{Han:1998sg}. The default choice for $\Lambda$ would be the fundamental scale $M_s$
unless mentioned otherwise.

Next we give the interference of SM $q\qb$ process with the gravity
mediated $q\qb$ subprocess. For convenience we will denote 
$M^{(0)}_{q\qb,sm} {M^{(0) *}_{q\qb,gr}} +c.~c.$ by  
$\Msq_{q \qb, int}$. 
\bea
\label{Mintlo}
\Msq_{q \qb, int} &=& \frac{e^2~ \kappa^2}{N}  
                    \left( C_0^q Z_0^q + C_1^q Z_1^q + C_2^q Z_2^q\right)  
\eea
\bea
\label{coupint}
Z_0^u  &=& \Gamma_z~ m_z~  C_v~ {\cal T}_Z~ \cot\theta_W ~\ImDs 
           \frac{1}{(s-m_z^2)^2+ \Gamma_z^2~m_z^2 } ~, 
\nonumber \\ [2ex]
Z_1^u  &=& {\cal T}_W^2~ \ReDs   ~, 
\\ [2ex]
Z_2^u  &=& \ReDs~ \Big[ C_v~ {\cal T}_Z ~\cot\theta_W
                   \frac{(s-m_z^2)} {(s-m_z^2)^2+ \Gamma_z^2~ m_z^2} 
                   + \frac{2~ Q_u}{s} \Big]      ~. 
\nonumber 
\eea
where $Z_i^q$ are in their essence combinations of EW couplings and 
propagator factors while $C_i^q$ are the functions of 
kinematic invariants. These $C_i^q$ functions are given below
\bea
\label{funcc}
C_1^u  &=&         \frac{1}{8 m^2 t}
           \Big[4m^8\left(-1 + n\right) + t^2\left(t - u\right)u - 
            m^6\left\{\left(-20 + 11n\right)t + nu\right\} + 
\nonumber \\ [2ex]
&&
            m^4t\left\{\left(-17 + 8n\right)t + \left(-11 + 4n\right)u\right\} - 
            m^2t\Big\{2\left(-2 + n\right)t^2 
\nonumber \\ [2ex]
&&
            + \left(-4 + n\right)tu + \left(-4 + n\right)u^2\Big\} \Big] ~,
\\ [2ex]
C_2^u   &=&       \frac{1}{8 m^2}
              \Big[-10m^6\left(-2 + n\right)\left(t - u\right) + m^4\left(-17 + 8n\right)\left(t^2 - u^2\right) + 
\nonumber \\ [2ex]
&&
            tu\left(t^2 - u^2\right) - 2m^2\left(-2 + n\right)\left(t^3 - u^3\right)\Big] ~,
\\ [2ex]
C_0^u  &=&        -C_2^u ~.
\eea
The $C_i^q$ expressions for down type quarks are related to that of up type quarks
as follows.
\bea
C_1^d \left(t,u,s \right) &=& C_1^u \left(u,t,s \right) ~,
\nonumber \\ [2ex]
C_2^d \left(t,u,s \right) &=& C_2^u \left(t,u,s \right) ~, 
\\ [2ex]
C_0^d \left(t,u,s \right) &=& C_0^u \left(t,u,s \right) ~. 
\nonumber
\eea

Note that the $W$ boson
polarization sum $-g_{\mu \nu} + k_\mu k_\nu / m^2$, which correctly
takes into account 3-polarizations of a massive particle, does not give rise to 
negative powers of $m$ and the $m \rightarrow 0$ limit is smooth. 

\subsection{Radiative Corrections}

In Fig.~\ref{smvrt}, the order $a_s$ loop diagrams that appear in
SM and in Fig.~\ref{bsmvrt} the diagrams with a graviton propagator are presented. 
We will use Feynman gauge $\xi=1$ in the QCD sector and we retain the
term proportional to $\xi$ in the gluon-gluon-graviton vertex. However this term
does not contribute to the matrix element squares.
Here we consider only 5 flavors of quarks and treat them as massless.
These diagrams contribute through their interference with the leading order
diagrams. In general loop diagrams give ultraviolet divergences and infrared
divergences when the integration over loop momenta is carried out. 
We use dimensional regularization $(n=4+\epsilon)$ to regulate these divergences;
these divergences then appear as poles in $\epsilon$. Note however that
owing to the gauge invariance and the fact that the $KK$ gravitons couple to 
SM energy momentum tensor, a conserved quantity, this process is UV finite.
The external parton leg corrections vanish in 
dimensional regularization for massless partons.
From the loop Feynman diagrams in the appendix we find that all
2-,3-and 4-point loop integrals appear in the calculation.
The maximum rank of tensor integrals is 3 and originate from the fermion
box. These tensor integrals  were reduced to scalar integrals following 
the procedure
of Passarino-Veltman \cite{Passarino:1978jh}. The 4-point scalar integrals 
that appear in the 
$gg$ initiated {\it box} diagrams were taken from 
\cite{Duplancic:2000sk}.
The one loop matrix elements are recorded below. The finite pieces of matrix element
squares denoted by a superscript {\it fin} are given in the appendix. \\
The SM contribution is found to be
\bea
\overline{|M^{V}|^2}_{q \qb,sm}
  &=&
       a_s(\mu_R^2) f(\e,\mu_R^2,s) C_F ~\Bigg[~
       \Upsilon \left(\epsilon \right)
      ~
            {\Msq}_{q \qb, sm}
          + \overline{|M^{V}|^2}^{fin}_{q \qb,sm}
                    \Bigg],
\label{sm1}
\eea
the interference of SM with the gravity mediated processes are
\bea
\overline{|M^{V}|^2}_{q \qb,int}
  &=&
       a_s(\mu_R^2) f(\e,\mu_R^2,s) C_F ~\Bigg[~
       \Upsilon \left({\e} \right)
           {\Msq}_{q \qb, int} + \overline{|M^{V}|^2}^{fin}_{q \qb,int}
 \Bigg]
\\[2ex]
\overline{|M^{V}|^2}_{gg,int}
  &=&
       a_s(\mu_R^2){C_A }  ~\Bigg[~
            \overline{|M^{V}|^2}^{fin}_{gg,int}   \Bigg]\,,
\label{sm2}
\eea
Note that in the above $gg$ initiated SM diagrams with a quark triangle and
$\gamma/Z$ propagator do not contribute as they vanish in massless limit
due to Furry's theorem and weak isospin invariance. Also note that 
$ \overline{|M^{V}|^2}_{gg,int} $ is completely finite; it does not contain
any soft or collinear divergences because in SM the $gg$ contribution 
begin at the loop level and a lowest order term should be finite. 

The pure gravity contributions are
\bea
\overline{|M^{V}|^2}_{q \qb,gr}
  &=&
       a_s(\mu_R^2) f(\e,\mu_R^2,s) C_F ~\Bigg[~
       \Upsilon \left({\e} \right)\Msq_{q\qb, gr}
      +  4(2\zeta(2) -5)  \Msq_{q\qb, gr}
       ~ \Bigg]
\\ [2ex]
\overline{|M^{V}|^2}_{gg,gr}
  &=&
       a_s(\mu_R^2) f(\e,\mu_R^2,s) C_A ~\Bigg[~
 \left\{-\frac{16}{\e^2} + \frac{4}{C_A \e}  \left( {11 \over 3 } C_A -\frac{4}{3}n_f T_f\right)\right\}
{\Msq}_{gg, gr}
\nonumber \\[1ex]
              &+& \frac{1}{9} \left( 72 \zeta(2) + 70 \frac{n_f T_f}{C_A} -203  \right)
{\Msq}_{gg, gr}
\Bigg]
\eea
where
\bea
\Upsilon \left({\e} \right)
         &=& -~\frac{16}{\displaystyle{\e^2}} + \frac{12}{\displaystyle{\e}}  ,\quad \quad \quad
f(\e,\mu_R^2,s)  = \frac{ \Gamma\left( 1+{\displaystyle {\e\over 2}}\right)} {\Gamma(1+\e)}
             \left( \frac{s}{4 \pi \mu_R^2} \right)^{\frac{\displaystyle{\e}}{2}}
\eea
The theory is renormalized at scale $\mu_R$. $C_F$ is the Casimir of the fundamental representation
while $C_A$ is the Casimir of adjoint representation in the color group.
\bea
\label{casimir}
C_F &=& \frac{N^2-1}{2N} , \quad \quad C_A = N ,\quad \quad T_f = \frac{1}{2} 
\eea
We can now write the 
order $a_s(\mu_R^2)$ contributions coming from virtual diagrams as,
%
\bea
\label{xvirt}
d \s^{virt} &=& a_s(\mur^2) dx_1dx_2 f(\e, \mur^2,s)
\nonumber \\[2ex]
           && \times
              \Bigg[ C_F \left(-\frac{16}{\e^2} + \frac{12}{\e} \right)
              \sum_i  d\s^{(0)}_{q_i\qb_i}(x_1,x_2,\e)  \left( f_{q_i}(x_1)f_{\qb_i}(x_2) + x_1 \leftrightarrow x_2 \right)
\nonumber \\[2ex]
           &&    + C_A \left\{-\frac{16}{\e^2} + \frac{4}{C_A\e} \left( \frac{11}{3}C_A -\frac{4}{3}n_fT_F \right) \right
\}d \s^{(0)}_{gg}(x_1,x_2,\e) \Big(f_g(x_1)f_g(x_2) \Big)
\nonumber \\[2ex]
&& + C_F \sum_i  d\s^{V,fin}_{q_i\qb_i}(x_1,x_2,\e)  \left( f_{q_i}(x_1)f_{\qb_i}(x_2) + x_1 \leftrightarrow x_2 \right)
\nonumber \\[2ex]
&& + C_A ~d\s^{V,fin}_{gg}(x_1,x_2,\e) (f_g(x_1)f_g(x_2))
\Bigg] ~.
\eea
The poles of order 2 in $\epsilon$ in the one loop matrix elements 
correspond to the configurations which are both soft and collinear simultaneously.
These double poles cancel when real emission contributions are included, the
remaining simple poles do not cancel completely and are factorized into the 
bare parton distribution functions at the scale $\mu_F$.

Several checks ensure the correctness of the matrix elements. 
The $W$ boson
polarization sum $-g_{\mu \nu} + k_\mu k_\nu / m^2$, which correctly
takes into account 3-polarizations of a massive particle, does not give rise to 
negative powers of $m$ and the $m \rightarrow 0$ limit is smooth. 
Further, for gluon initiated process the gluon polarization sum is 
$-g_{\mu \nu} + (k_\mu n_\nu + k_\nu n_\mu)/k.n$ where $n$ is an
arbitrary light like vector and the results are independent of the vector
$n$.  Furthermore the SM matrix elements are in agreement with the literature
\cite{Ohnemus:1991kk}.
 
\begin{figure}[ht]
\centerline{\epsfig{file=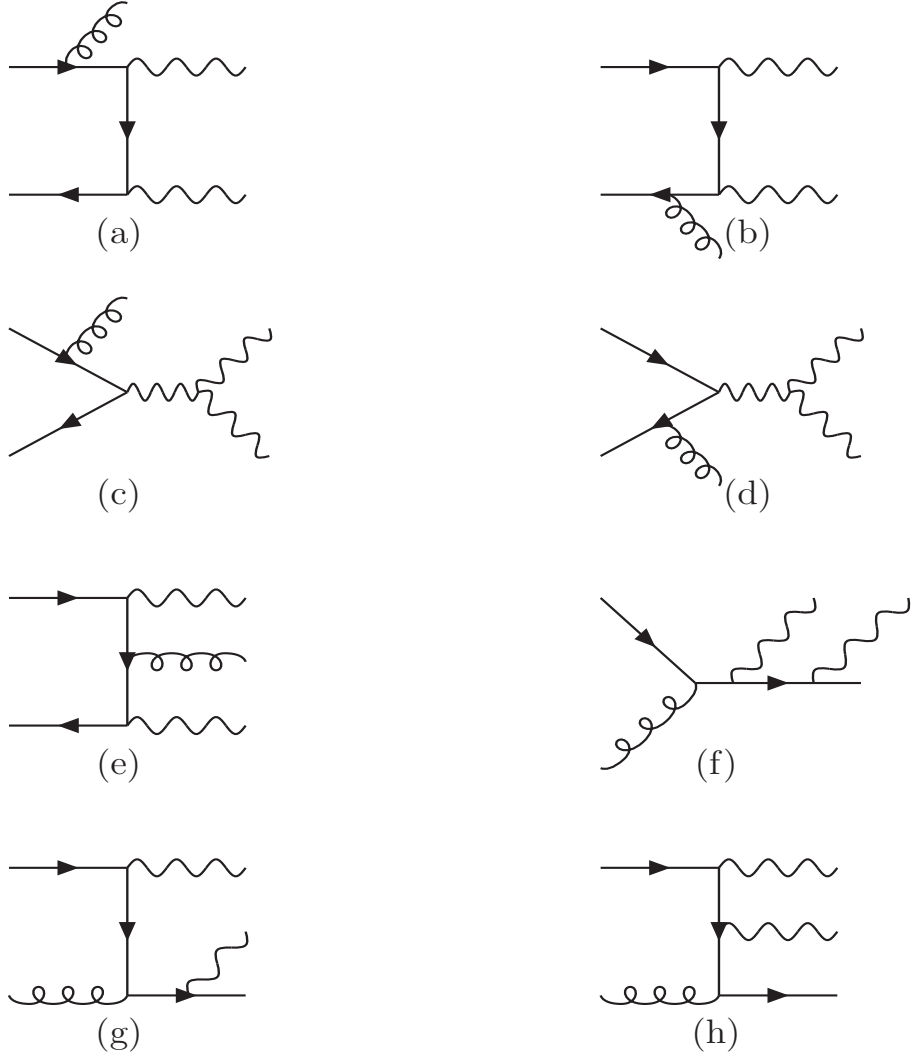,width=12cm,height=14cm,angle=0} }
\caption{Order $a_s$ real emission Feynman diagrams in SM for the subprocess
  $ u\bar{u}\rightarrow W^+W^-g$ and $u g \rightarrow W^+W^-u$. 
  The diagrams for the subprocess 
  $ d\bar{d}\rightarrow W^+W^-g$ and $ d g \rightarrow W^+W^- d$
  are obtained by replacing $u \rightarrow d$
  and $W^+ \leftrightarrow W^-$ in the diagrams shown here.}
\label{smreal}
\end{figure}

\begin{figure}[ht]
\centerline{\epsfig{file=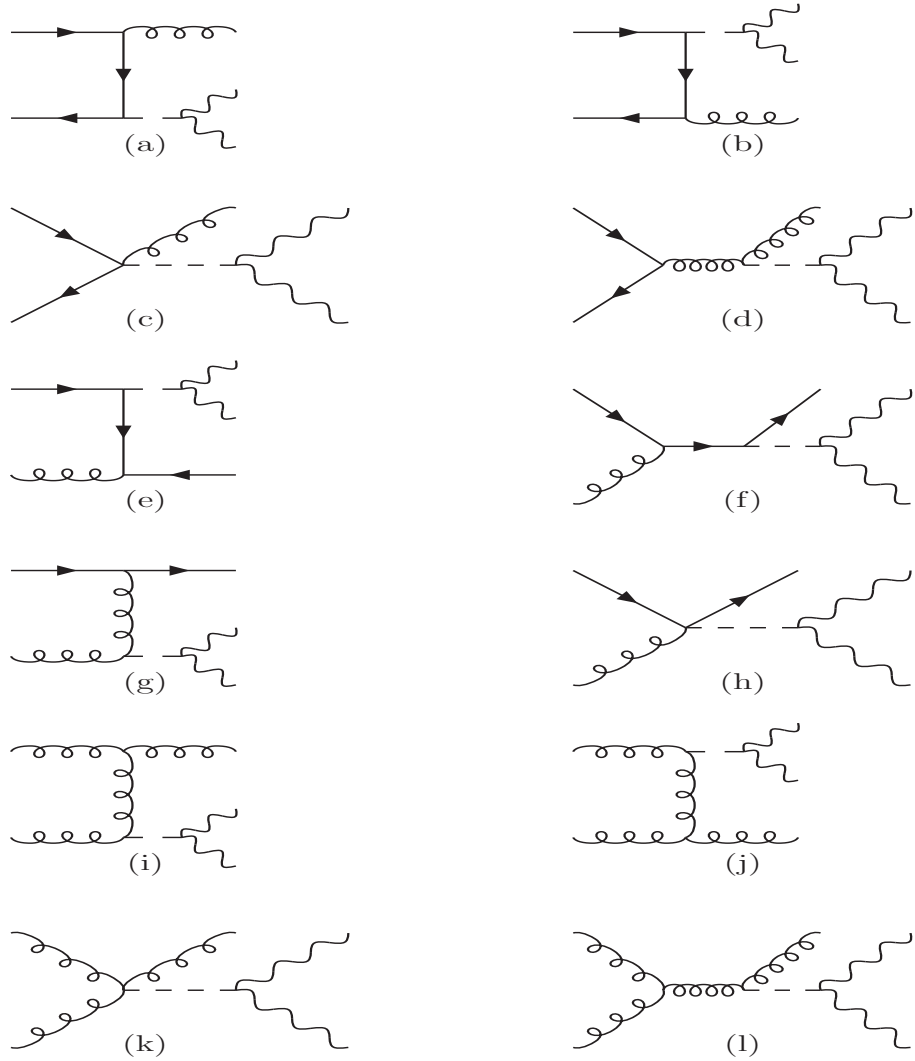,width=12cm,height=14cm,angle=0} }
\caption{Gravity mediated real emission diagrams for the subprocess
  $ q\bar{q}\rightarrow W^+W^-g$, $q g \rightarrow W^+W^-u$ and
  $ gg \rightarrow W^+W^-g$. 
The big dashed
lines represent gravitons.} 
\label{bsmreal}
\end{figure}

At NLO we also have to include $2 \rarrow 3$ real emission processes.
A generic process is of the form
\be
a(p_1) + b(p_2) \rarrow W^{+}(p_3) + W^{-}(p_4) + c(p_5).
\ee
In Fig.~\ref{smreal} we show the $q \qb$ and $qg$ initiated real emission Feynman diagrams
which appear in SM. In addition, in the ADD model the $2 \rarrow 3$ diagrams
with graviton propagator are shown in Fig.~\ref{bsmreal}. Here all the three kinds,
$q\qb, qg, gg$ initiated subprocesses occur. The $2 \rarrow 3$ contributions to 
cross-section reveal the infrared divergences when the integral over the 
final state particles is carried out. The sum of virtual and
real emission cross section is finite after mass factorization is carried out.
For details we refer to  the review \cite{Harris:2001sx}.

Although the details of phase space slicing method to deal with soft and collinear
singularities in real emission processes were given
in our earlier works \cite{Kumar:2008pk,Agarwal:2009xr},
we shall recapitulate it for completeness. 
The $2 \rarrow 3$ phase space is divided into soft and collinear regions
using two small dimensionless slicing parameters $\delta_s$ and $\delta_c$ .
The soft region is defined as the part of phase space where the final state gluon is soft and 
has an energy less than $\delta_s {\sqrt s_{12}}/2$ in the center of mass frame 
of incoming partons.
The region complementary to the soft region is hard region and contains 
collinear singularities. This region is thus further divided into hard 
collinear region (the region of phase space where the final state parton
is collinear to one of the initial state parton) which contains collinear
singularities and hard non-collinear region which is free of any singularities.
All the order $a_s$ pieces together; the virtual cross-section $d\sigma^{virt}$
the soft piece $d\sigma^{soft}$ 
and the mass factorized hard collinear contribution  
$d\sigma^{HC+CT}$ (CT denotes mass factorization counter term) is referred to as 2-body contribution.
\be
d\sigma^{2-body}(\ds,\dc,\mu_F) = d\sigma^{virt}+d\sigma^{soft}(\ds,\dc)+d\sigma^{HC+CT}(\ds,\dc,\mu_F).
\ee
The only order $a_s$ piece, $d\sigma^{3-body}(\ds,\dc)$, 
which remains to be included is hard non collinear and which is finite 
as the integration over 3-body phase space
here does not include soft and collinear regions. 
The integration over the 3-body phase space is carried out using monte carlo,
and it is constrained to avoid collinear and  soft regions. The $q\qb$ and $gg$
initiated processes contain both kinds of divergences so the integral
is constrained using $\ds$ and $\dc$ to avoid these regions. The $qg$ initiated 
process, however, contain only collinear singularities (as soft fermions do not
give singularities) and the 3-body integration is constrained using only $\dc$.
 
The NLO result is sum $d\sigma^{LO} + d\sigma^{2-body}(\ds,\dc,\mu_F) + d\sigma^{3-body}(\ds,\dc)$.
The sum $d\sigma^{2-body}$ $(\ds,\dc,\mu_F)$ $+ d\sigma^{3-body}(\ds,\dc)$ constitutes QCD correction,
but  $d\sigma^{2-body}(\ds,\dc,\mu_F) $ and \\
$ d\sigma^{3-body}(\ds,\dc)$ independently are not
physical quantities as these depend on the (arbitrary) slicing parameters. The sum of these two 
pieces should be independent of the slicing parameters as these were introduced at the intermediate
stages of calculation. 
We have checked that the sum of 2-body and 3-body contribution is independent
of the $\delta_s$ and $\delta_c$ over a large range of their values.

In the next section we present the results using our monte carlo code
which incorporates the above given details. 
We will present the stability
of results against variations of the slicing parameters.
This code can easily 
accommodate any cuts on the final state bosons and can evaluate 
various kinematical distributions. 

\section{Kinematical distributions and Results}
The LHC with a center of mass energy of $14~TeV$ will be our 
default choice. However we will also present some results for a center of mass energy of $10~TeV$ for the LHC.
For numerical evaluation,
the following SM parameters 
\cite{Amsler:2008zzb} are used
\be 
m_W= 80.398~ GeV,  \quad m_Z = 91.1876~ GeV,\quad \Gamma_Z=2.4952 ~GeV, \quad \sin^2 \theta_W = 0.231
\ee
where $\theta_W$ is the weak mixing angle.
For the electromagnetic coupling constant $\alpha$ we use $ \alpha^{-1} = 128.89$. 
CTEQ6 \cite{Pumplin:2002vw} density sets are used for parton distribution 
functions. 2-loop running for the strong coupling constant is used .
The number of active massless-quark flavors is taken to be 5 and the value of $\Lambda_{QCD}$ is 
chosen as prescribed by the CTEQ6 density sets. At leading order, that is at order $\alpha_s^0$,  we
use CTEQ6L1 density set (which uses the LO running $a_s$) with the corresponding 
$\Lambda_{QCD}=165~MeV $. At NLO we use CTEQ6M density set ( which uses 2-loop running $a_s$ )
with the $\Lambda_{QCD}=226~MeV $; this value of $\Lambda_{QCD}$ enters into the evaluation of the 
2-loop strong coupling.
The  default choice for the renormalization and factorization scale is the identification
to the invariant mass of the $W$ boson pair ie., $\mu_F =\mu_R =Q$. Furthermore the
$W$ bosons will be constrained to satisfy $|y_W| < 2.5$, where $y_W$ is 
the rapidity of a final state $W$ boson .

\begin{figure}[ht]
\centerline{\epsfig{file=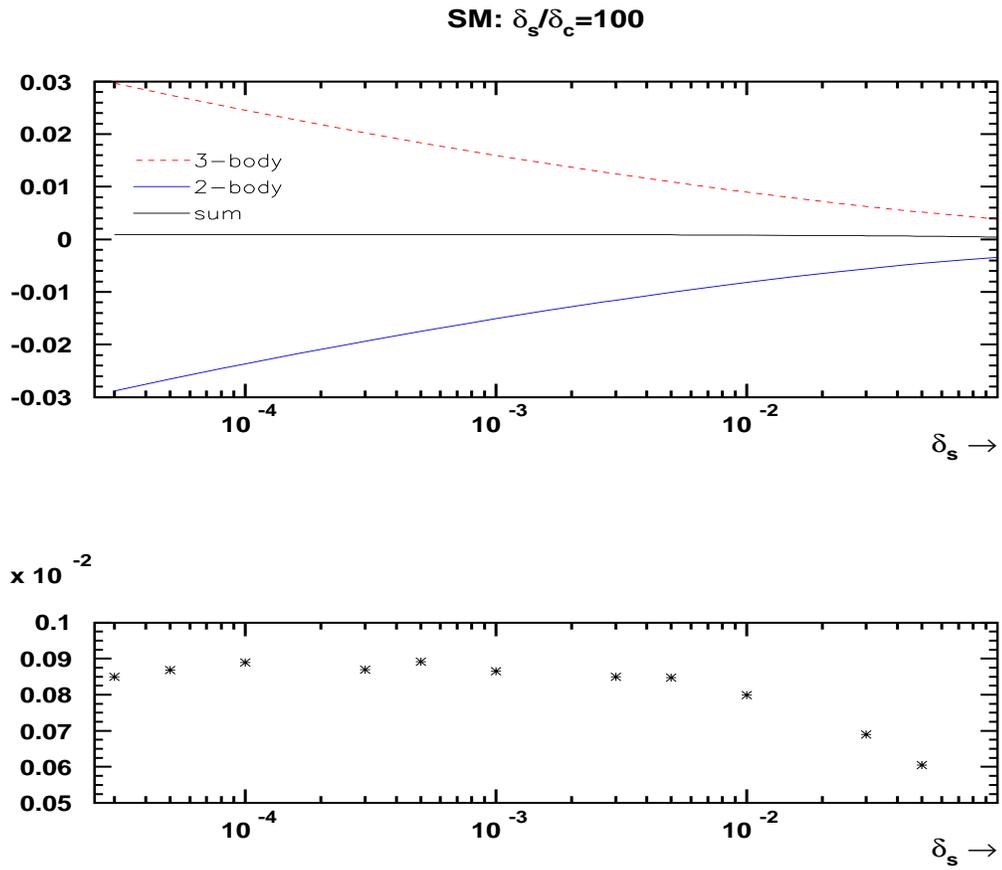,width=15cm,height=12cm,angle=0} }
\caption{Variation of 2-body and 3-body contributions (of $d\sigma /dQ$ at $Q=800~GeV$ in SM) and their sum
with $\delta_s$.
Here $\delta_s/\delta_c =100$ has been used.  }
\label{dltsm}
\end{figure}

\begin{figure}[ht]
\centerline{
\epsfig{file=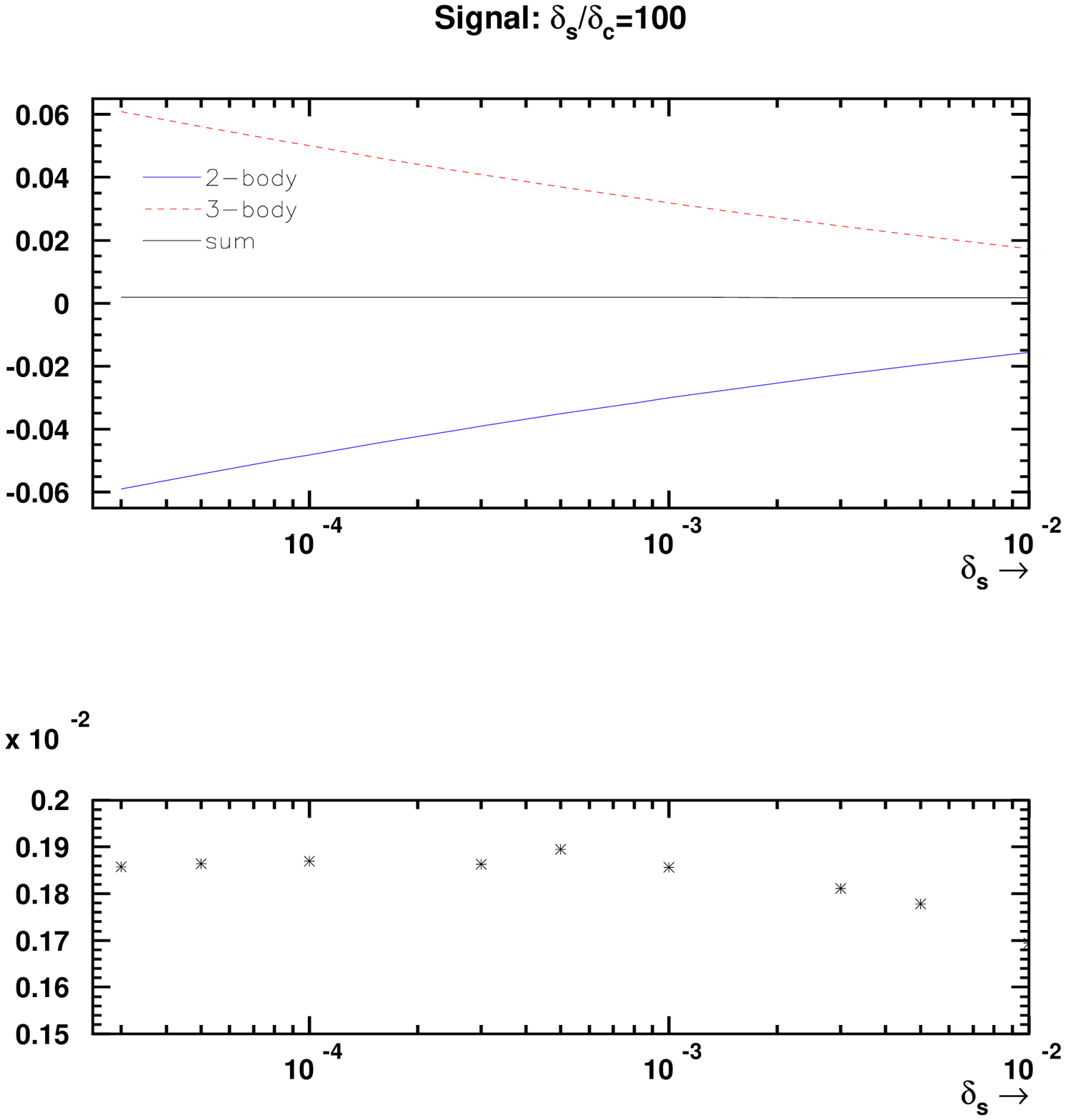,width=15cm,height=12cm,angle=0}}
\caption{Variation of 2-body and 3-body contributions (of $d\sigma /dQ$ at $Q=800 GeV$ in signal) 
and their sum with $\delta_s$.
Here $\delta_s/\delta_c =100$ has been used.  }
\label{dltsig}
\end{figure}

We will present below the following kinematical distributions: 
\begin{enumerate}
\item
Invariant mass 
distribution, $d\sigma /dQ$, where $Q$ is the invariant mass of the
final state $W$ boson pair, 
\item 
Rapidity distribution $d \sigma/ dY$ 
where 
$Y =  1/2~ \ln (P_1\cdot q)/(P_2 \cdot q) $,
where $P_1$ and $P_2$ are incoming proton momenta and $q$ is the  sum
of the $W$ boson 4-momenta. 
\end{enumerate}
First we demonstrate that the sum of 2-body and
3-body contributions is fairly independent of the slicing parameters. In Fig.~\ref{dltsm} (for SM) and 
Fig.~\ref{dltsig} (for signal) we show the variations of these two pieces with the slicing parameters in 
invariant mass distribution at a value of invariant mass 
equal to $800~ GeV$. Here both $\ds$ and $\dc$
are varied together with the ratio $\ds/\dc$ fixed at a value of 100 
\cite{Harris:2001sx}. We note that the sum of 
2-body and 3-body contributions is fairly stable against variations in these 
parameters and this gives us confidence in our code.
In what follows we will use $\delta_s =10^{-3}$ and $\delta_c=10^{-5}$.

\begin{figure}[ht]
\centerline{\epsfig{file=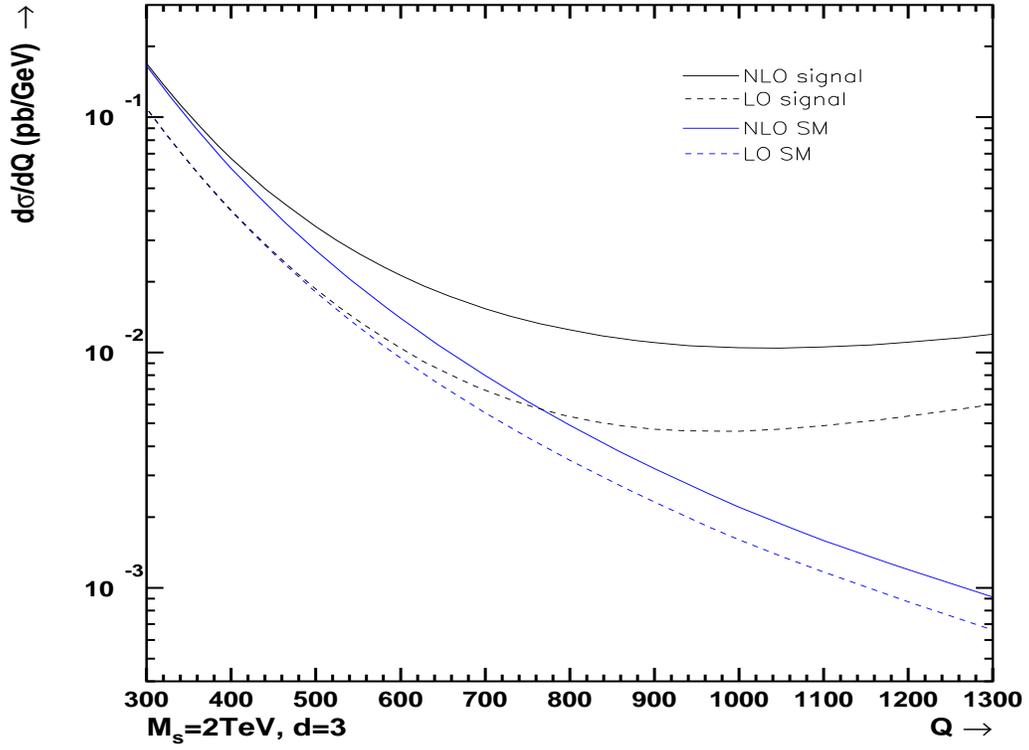,width=15cm,height=12cm,angle=0} }
\caption{Invariant mass distribution at LO and NLO in SM and for the signal at $M_s=2TeV$ and 
3 extra dimensions. }
\label{inv}
\end{figure}

In Fig.~\ref{inv} we have plotted the invariant mass distribution both for the SM and the 
signal, in the range 300 $GeV$ to 1300 $GeV$. 
In this plot we display
for three extra dimensions ie., $d=3$ and for fundamental scale equal to $2~TeV$. 
To highlight the importance of
QCD corrections we have also displayed the LO results of SM and the signal, and 
we observe that the $K$ factors (defined as $K=d\sigma^{NLO}/d\sigma^{LO}$) 
are significantly large. 
We note that for the signal ($sm+gr+int$) $K$ factor varies between 1.55 to 1.98 
in the invariant mass range of 300 to 1300 $GeV$. 
This also shows that the LO results can be only 
treated as first approximations and to have more precise estimates we should
go beyond the leading order. We note here that present computation does not 
take into account decay of $W$ bosons to leptons which is observed 
experimentally, but as QCD corrections are 
independent of these decays, the $K$ factors obtained here would not change
when decays are taken into account.

\begin{figure}[ht]
\centerline{\epsfig{file=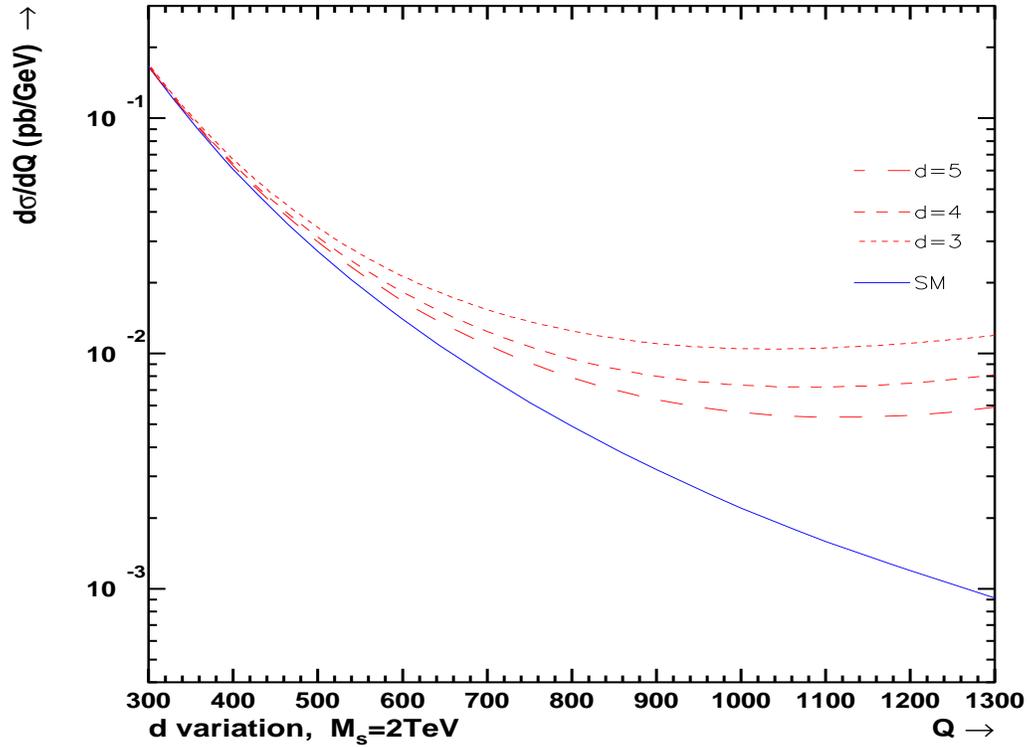,width=15cm,height=12cm,angle=0} }
\caption{
Effect of variation of number of extra dimensions in invariant mass distribution.
The fundamental scale $M_s$ has been fixed at 2 TeV.
The curves correspond to NLO results.  }
\label{invd}
\end{figure}

\begin{figure}[ht]
\centerline{\epsfig{file=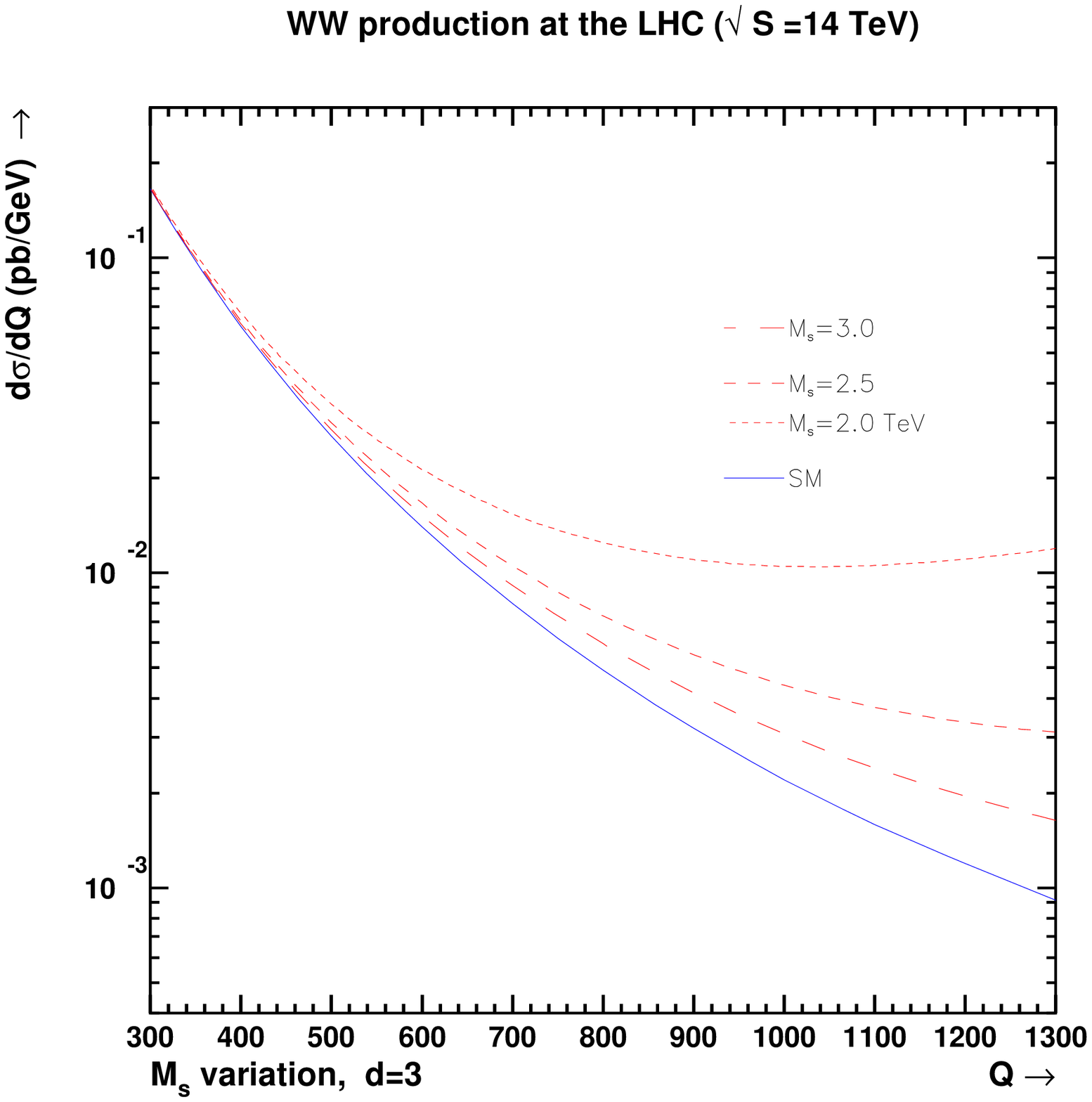,width=15cm,height=12cm,angle=0} }
\caption{Effect of variation of the fundamental scale $M_s$ in the invariant mass
distribution. The number of extra dimensions has been fixed at 3. The curves correspond
to NLO results.
}
\label{invm}
\end{figure}

\begin{figure}[ht]
\centerline{\epsfig{file=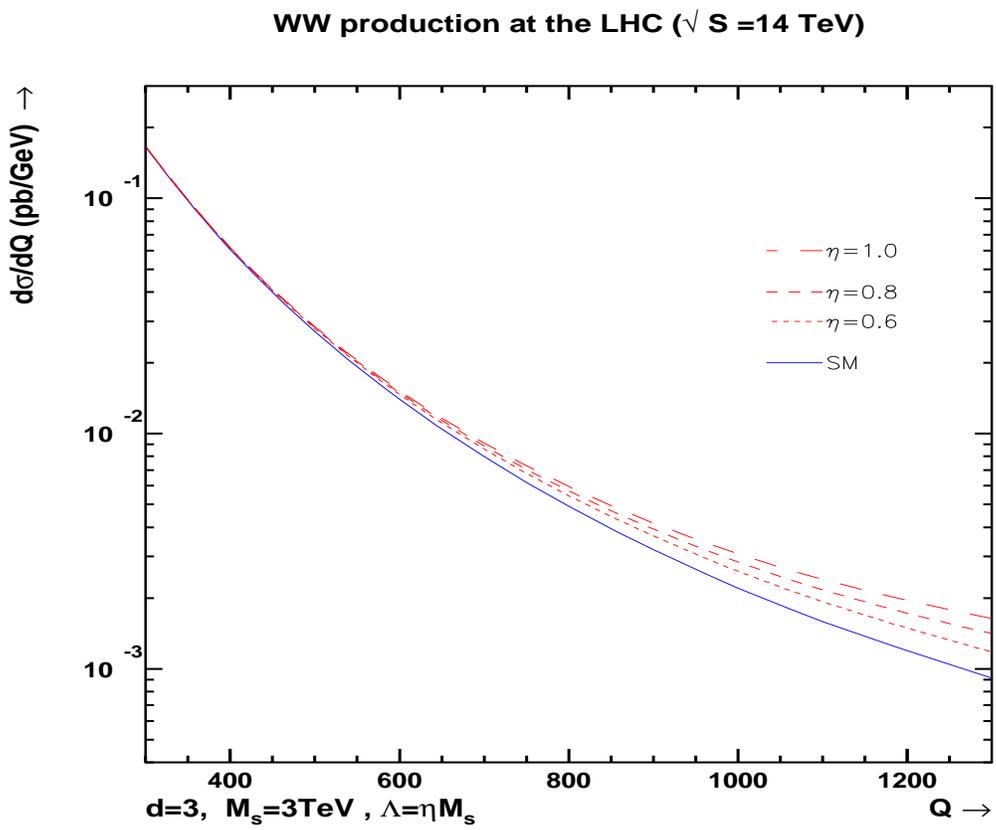,width=15cm,height=12cm,angle=0} }
\caption{Effect of variation of the UV cutoff scale $\Lambda$ in the invariant mass
distribution for $d=3$ and $M_s =3 TeV$. The curves correspond
to NLO results.
}
\label{UV}
\end{figure}

To estimate the effect of the number of extra dimension on the invariant mass
distribution, we plot in Fig.~\ref{invd} the signal for three different values of $d$ (3,4,5)
with $M_s$ fixed at 2 $TeV$. We note that the lower the value of $d$, more is 
the strength of the signal. Next in Fig.~\ref{invm} we have plotted $d\sigma/dQ$ for 
three different values of $M_s$ (2.0, 2.5, 3.0) at a fixed value 3 for the  number of
extra dimensions. As expected, with increase in the fundamental scale the deviations
from SM predictions become less, and significant deviations from SM are observed 
at higher energies still. Next, in Fig.~\ref{UV}  we present the effect of variation of the UV cutoff $\Lambda$
that appears in the expression of graviton propagator (see eq. \ref{prop}) for $M_s=3TeV$ and $d=3$. We 
observe that lowering the cutoff from $M_s$ by 25\% lowers the predictions by 14\%. 

\begin{figure}[ht]
\centerline{\epsfig{file=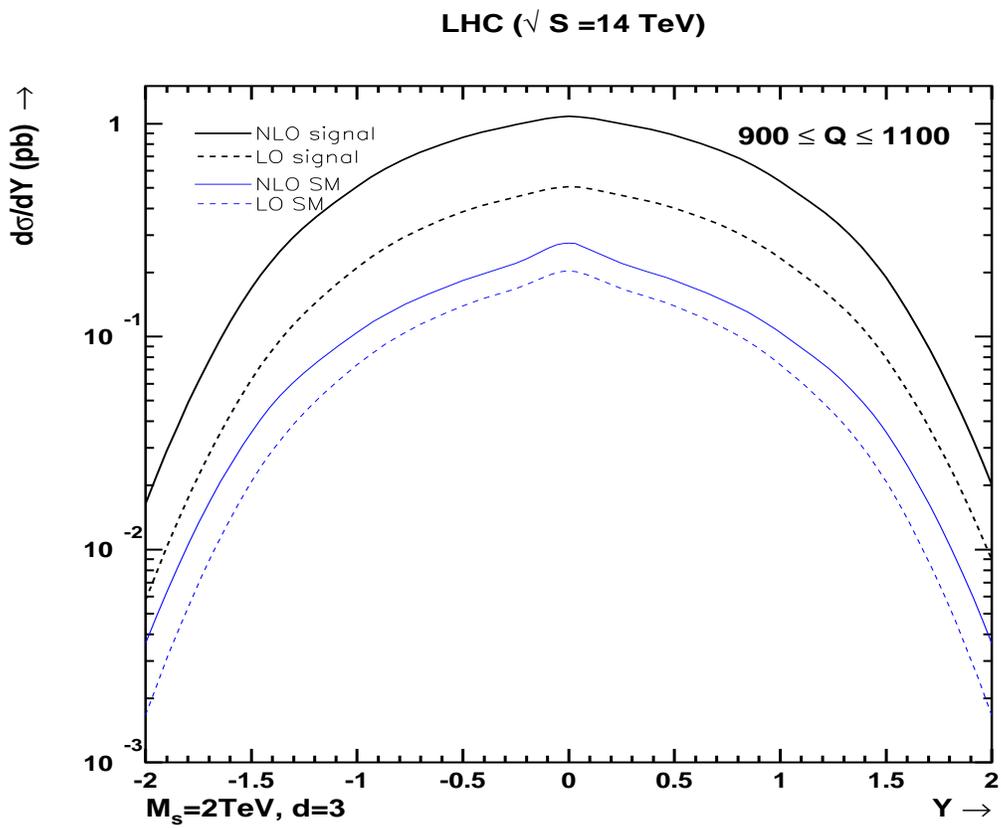,width=15cm,height=12cm,angle=0} }
\caption{
Rapidity distribution for $M_s =2TeV$ for SM and signal for $d=3$ . 
We have integrated over the 
invariant mass range $900<Q<1100$ to enhance the signal.
}
\label{Y}
\end{figure}

In Fig.~\ref{Y} we have plotted the rapidity distribution $d\sigma/dY $ at LO and 
NLO both for SM and the signal for $d=3$ and $M_s$ fixed at 2 TeV. 
We have plotted this distribution in the interval $-2.0 < Y < 2.0 $ and 
have carried out an integration over the invariant mass interval $900 < Q < 1100$
to increase the signal over the SM background.
As expected the distribution is symmetric about $Y=0$.
 
\begin{figure}[ht]
\centerline{\epsfig{file=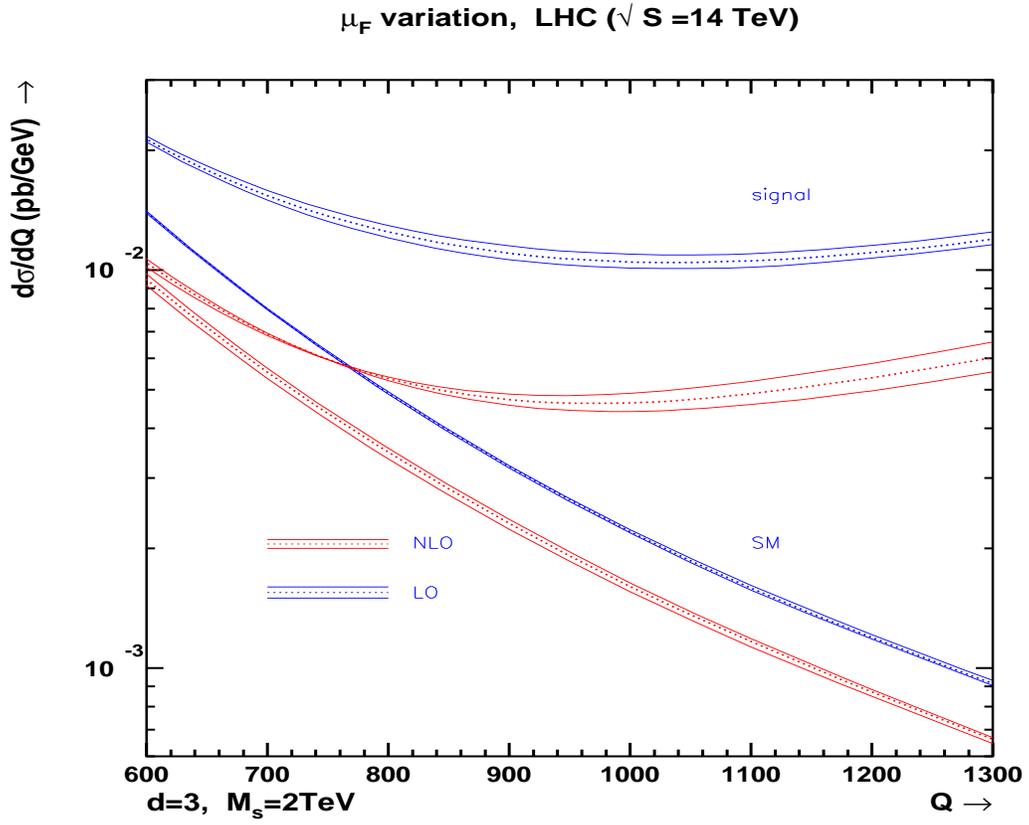,width=15cm,height=12cm,angle=0} }
\caption{Factorization scale variation in the invariant mass distribution. The number of 
extra dimensions $d=3$ and the fundamental scale $M_s=2TeV$ have been chosen. 
}
\label{mufvar}
\end{figure}

\begin{figure}[ht]
\centerline{\epsfig{file=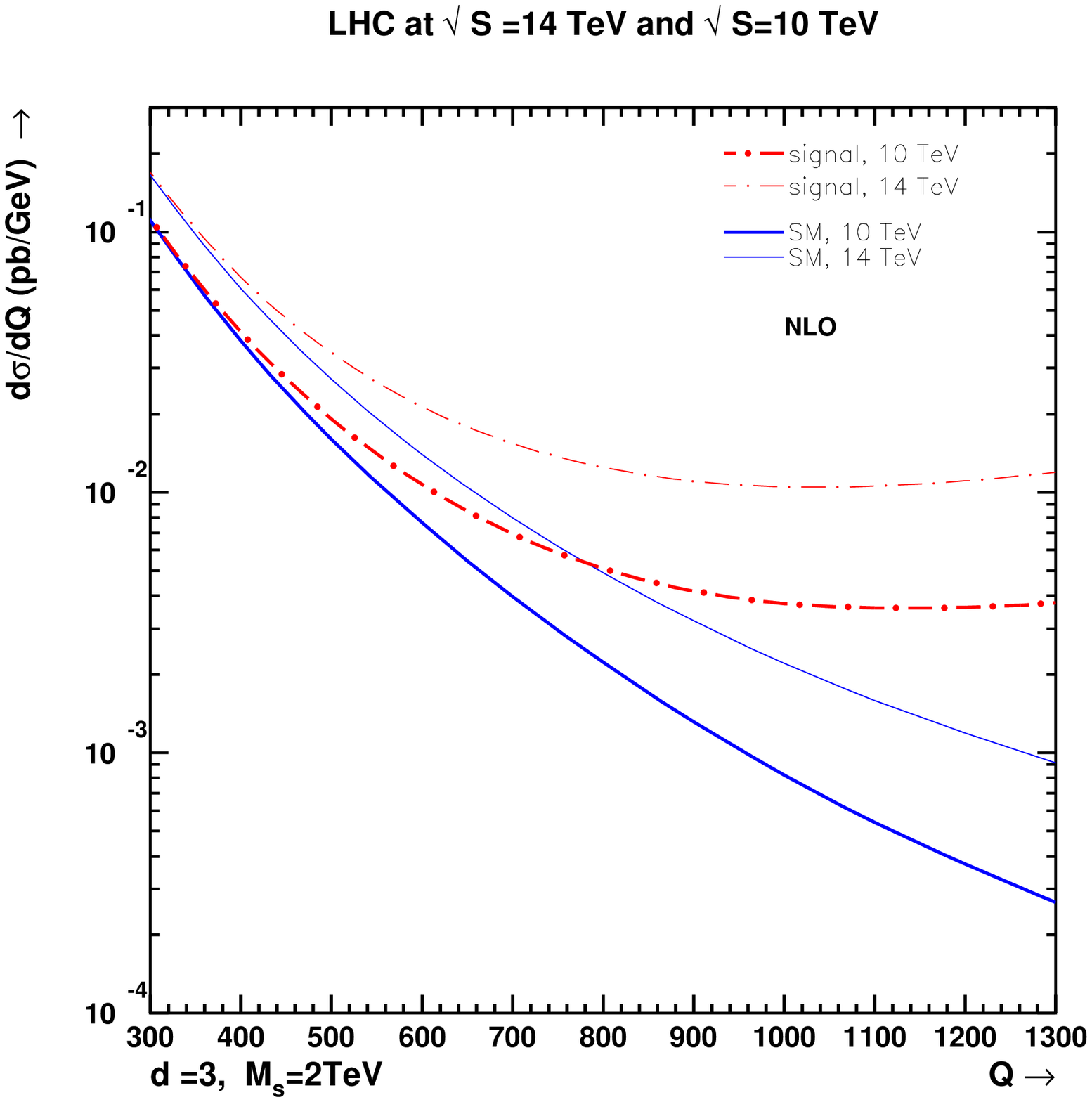,width=15cm,height=12cm,angle=0} }
\caption{
Invariant mass distribution at NLO for SM and the signal. Here the thicker
curves correspond to $\sqrt S =10 TeV$ and lighter curves to $\sqrt S=14 TeV$
at the LHC.
}
\label{ten}
\end{figure}

As was noted before the NLO QCD corrections reduce the sensitivity 
of the cross sections to the factorization scale $\mu_F$; this we now show in 
the Fig.~\ref{mufvar}. We have plotted SM and the signal both at LO and NLO,
and have varied the factorization scale $\mu_F$ in the range $Q/2 < \mu_F < 2Q$.
The central curve in a given band (shown by the dotted curves) correspond to 
$\mu_F =Q$. In all these the renormalization scale is fixed at $\mu_R =Q$.
We notice that the factorization scale uncertainties in SM are less compared
to the signal. 
This is because of the dominant role of the gluon gluon initiated
process in the signal. 
We see in this figure that a significant
reduction in theoretical uncertainty, arising from the factorization scale, is achieved 
by our NLO computation. At $Q=1300~GeV$ the $d\sigma /dQ$ for the signal varies by 18.8 \% at LO
as $\mu_F$ is varied between $Q/2$ to $2Q$ and it varies by 7.6 \% at NLO. 
At the end we present in Fig.~\ref{ten}, $d\sigma/dQ$ for LHC with a centre of mass
energy of $10~TeV$ at NLO both for SM and signal. For comparison 
we have also plotted the $14~TeV$ results in the same figure.

\section{Conclusions}
In this paper we undertook computation of $W$ boson pair production at the
LHC at next-to-leading order in QCD in the extra dimension model of ADD.
Here only spin-2 $KK$ gravitons appearing at the propagator level were
considered.
$W$ boson production is one of the important channels at the LHC to probe 
both the 
standard model and new physics. As the leading order results serve only
as first approximations we need to go beyond it to NLO to have more
precise estimates. The NLO results are generally not only significantly
larger as compared to the LO results but they are also much less 
sensitive to the arbitrary factorization scale and renormalization scale
(if the LO already starts at order $a_s$). 

Here we carried out a full NLO computation and 
presented analytical expressions of matrix element squares 
for all the SM, gravity mediated and the interference of SM and 
gravity mediated processes both at the 
LO and virtual level. We used dimensional regularization to 
regulate soft and collinear divergences and the singularity 
structure in the loop level matrix elements is shown and it
is observed that the singular pieces are proportional to the
born contributions. 
As different kinematical distributions such as invariant mass
distribution and rapidity distribution are evaluated with 
cuts on the final state $W$ bosons it is useful to use monte
carlo based semi analytical methods which allow to tailor 
code easily to these requirements. For this we used the two 
cutoff phase space slicing method to divide phase space in 
soft and collinear regions and filter out the singularities in 
the real emission contributions which appear on phase space integrations.
A brief discussion on this method was presented.  
To save space we omitted the real emission 
matrix elements as the expressions are voluminous and can be obtained
on request. We have used ${\overline MS}$ scheme throughout this paper.

We have presented distributions for 
the LHC at $14 TeV$ and $10 TeV$.
We first offered some checks on our monte carlo code such as 
stability of sum of $2-body$ and $3-body$ contributions against variation
of the slicing parameters $\delta_s$ and $\delta_c$ and then presented 
invariant mass and rapidity distributions both at LO and NLO. 
We use CTEQ 6L1 and CTEQ 6M parton density sets for LO and NLO observables, respectively.
Significant enhancements over the LO predictions are observed.
The $K$ factors are found to be large in the invariant mass distribution. For LHC at
$14 TeV$ we find that $K$ factor in the invariant mass distribution for the 
signal ($sm+gr+int$) varies between 1.55 and 1.98 as $Q$ varies between 300 to 1300 $GeV$.
We have shown that a significant
reduction in theoretical uncertainty, arising from the factorization scale, is achieved 
by our NLO computation. At $Q=1300~GeV$ the $d\sigma /dQ$ varies by 18.8 \% at LO
as $\mu_F$ is varied between $Q/2$ to $2Q$ and it varies by 7.6 \% at NLO. 
These observations justify the entire exercise and 
give results that are precise and  
suitable for further studies for constraining the parameters of the ADD model.
Invariant mass distribution is also presented for LHC at a center of mass energy of
$10 TeV$ at the NLO level.
\\

\noindent
{\bf Acknowledgments:}
The work of NA is supported by CSIR Senior Research Fellowship, New Delhi.
NA, AT and VR would also like to thank
the cluster computing facility at Harish-Chandra Research Institute. 
NA and VKT acknowledge the computational support of the computing facility which has been developed by
the Nuclear Particle Physics Group of the Physics Department, Allahabad University under the Center of
Advanced Study (CAS) funding of U.G.C. India.  The authors
would like to thank Prakash Mathews and M.C. Kumar for useful discussions.



\section{Appendix}
Below we give the finite pieces of the matrix element squares
that appear in Eq. \ref{sm1} through Eq. \ref{sm2} in section 2.
\bea
\overline{|M^{V}|^2}^{fin}_{q \qb,sm} 
  &=&  \frac{e^4}{N} \left[A_0^q B_0^q + A_1^q B_{11}^q +A_2^q B_{22}^q + A_3^q B_{33}^q\right]
\eea
where $A_i^q$ contain the coupling and propagator factors , $B_i^q$ are the functions of the 
kinematic invariants.
\be
A_0^u = 
        \frac{1} {2 \pi} 
\frac{ (C_a+C_v)~ \Gamma_z~ m_z~ {\cal T}_W^2~ {\cal T}_Z~ \cot\theta_W}
{ (s-m_z^2)^2+\Gamma_z^2m_z^2}
\ee
and the rest $A_i^q$ are given in eqn. (\ref{coup}).
The $B_i^q$ functions are as follows.
\bea
B_0^u &=&       \frac{12}{t}
             \Big[2G_1
            \left (10m^4 - 4m^2(2t + u) + t(2t + u) \right)  
\nonumber \\ [2ex] 
&& 
            -\frac{1}{(m^2-t)^2} \Big \{\zeta(2) \Big(15m^8 - 6m^6(2t + u) - t^2u(t + 2u) - 
                 3m^4t(t + 3u) 
\nonumber \\ [2ex] 
&& 
             + 2m^2t(t^2 + 6tu + 2u^2)
                \Big ) \Big \} \Big] ~,
\\ [4ex] 
B_{11}^u  &=&       \frac{2G_2}{t^2} \left (-2m^4 + tu \right) + 
               2\zeta(2) \left  \{7 - \frac{8m^4}{t^2} + \frac{6u}{t} 
               + \frac{tu}{m^4} - \frac{4(t + u)}{m^2} \right \}  
\nonumber \\ [2ex] 
&& 
          +\frac{G_3}{(m^2-t)t^2} \Big \{6m^6 + 6m^4t - 2t^2u 
           - 2m^2t(2t + 3u) \Big \}  
\nonumber \\ [2ex] 
&& 
          + \frac{2(2m^2 + t + u)}
             {m^4(-4m^2 + s)^2t^2} 
             \Big \{18m^{10} - 2t^3u(t + u) 
               + m^8(11t + 9u) - 
               2m^6t(14t + 9u)  
\nonumber \\ [2ex] 
&& 
             +  m^4t(2t^2 - 9tu - 9u^2) + 
               4m^2t^2(2t^2 + 3tu + 2u^2) \Big \}
           + \frac{G_4}{(-4m^2 + s)^2t^2} 
\nonumber \\ [2ex] 
&& 
           \Big \{24m^8 + 8m^6(t + 3u) + 
               2t(t - u)(t + u)(2t + 3u) + 
               4m^2t(t^2 - 6tu - 3u^2)  
\nonumber \\ [2ex] 
&& 
             +  m^4(-22t^2 - 8tu + 6u^2) \Big \}
               + \frac{2G_5}
                {s(-4m^2 + s)^2t }
            \Big \{-12m^8 + 4m^6(s + 2t) 
\nonumber \\ [2ex] 
&& 
             + su(t + u)^2 + 
               2m^2su(3t + u) + 2m^4(2t^2 + s(t + u)) \Big \} ~,
\eea
\bea
B_{22}^u &=&        \frac{1}{m^4}
            \Big[-4(-2 + \zeta(2))\Big\{40m^8 - 16m^6(t + u) - 
           tu(t + u)^2 + 4m^2(t + u)^3  
\nonumber \\ [2ex] 
&& 
            - m^4(7t^2 + 22tu + 7u^2)\Big\}\Big] ~, 
\\ [4ex] 
B_{33}^u &=&       2\Big[-\frac{G_2}{t}
             \Big  \{10m^4 - 4m^2(2t + u) 
           + t(2t + u) \Big \} + 
          \frac{2\zeta(2)}{m^4 t}
        \Big  \{-30m^8 
\nonumber \\ [2ex] 
&&          + t^2u(t + u) 
              - 4m^2t(t + u)^2 + 
               5m^4t(t + 2u) 
            + 4m^6(4t + 3u) \Big  \}  
\nonumber \\ [2ex] 
&& 
          +\frac{G_3}{t (m^2 - t)^2} 
          \Big  \{15m^8 - 6m^6(2t + u) 
            - t^2u(t + 2u) 
              - 3m^4t(t + 3u) 
\nonumber \\ [2ex] 
&& 
            + 2m^2t(t^2 + 6tu + 2u^2) \Big \} + 
          \frac{G_5}{s(-4m^2 + s)^2 }
            \Big \{-32m^8 + 4m^6(11s + 8t)  
\nonumber \\ [2ex] 
&& 
              - s(t + u)^2(2t + u) 
              + 2m^4s(5t + 17u) + 
               2m^2s(-2t^2 + tu + 3u^2) \Big \}  
\nonumber \\ [2ex] 
&& 
         +\frac{G_4(2m^2 + t + u)}{t (-4m^2 + s)^2}
           \Big  \{30m^6 + m^4(-13t + 3u) 
              + t(t + u)(2t + 3u)  
\nonumber \\ [2ex] 
&& 
           -    2m^2(t^2 + 11tu + 3u^2) \Big  \}
           +\frac{ (2m^2 + t + u)^2}
           {m^4(-4m^2 + s)^2(m^2 - t)t}
            \Big \{85m^{10} 
\nonumber \\ [2ex] 
&& 
              + m^6t(4t - 17u) 
              + 4t^3u(t + u) - 
               4m^2t^2(t + u)(4t + 5u)  
\nonumber \\ [2ex] 
&& 
              - m^8(115t + 34u) + 
               m^4t(44t^2 + 79tu + 18u^2)
             \Big  \}
            \Big] ~. 
\eea

\pagebreak

\bea
\!\!\!\!\overline{|M^{V}|^2}^{fin}_{q \qb,int} 
  &=& = \frac{e^2 \kappa^2}{N} \left[C_{00}^q Z_0^q + C_{11}^q Z_1^q +C_{22}^q Z_2^q +C_{33}^q Z_3^q\right] 
\eea
where $Z_i^q$ contain the coupling and propagator factors while $C_i^q$ are the functions of 
kinematic invariants.
\be
Z_3^u=  {\cal T}_W^2 ~\pi ~\ImDs
\ee
and rest $Z_i^q$ are given in eqn. (\ref{coupint}).
The $C_i^q$ functions are as follows. 
\bea
C_{11}^u &=&        \frac{1}{4}\Big[ 
               \frac{2G_3}{t (m^2 - t) }
                \Big  \{-9m^8 - 4m^4t^2 + 3m^6(5t + u) - 
                     m^2tu(9t + u) + t^2u(2t + 3u)
                 \Big  \}
\nonumber \\ [2ex] 
&& 
              + 4G_8
               \Big  \{\frac{-3m^6}{t} + t^2 + m^2 \left(-5t 
                   - \frac{7u}{2} \right) + tu + 
                    \frac{u^2}{2} + m^4 \left (8 + \frac{u}{t}\right)
               \Big  \} 
\nonumber \\ [2ex] 
&& 
             - \frac{4G_6}{t}
               \Big  \{-6m^6 
+ 2m^4(8t + u) - m^2t(10t + 7u) 
                   + t(2t^2 + 2tu + u^2)
                \Big \}
\nonumber \\ [2ex] 
&& 
             + \frac{4\zeta(2)}{m^2 t}
               \Big  \{18m^8 + t^2(t - u)u - 
                 2m^6(20t + 3u) + m^4t(25t + 12u)  
\nonumber \\ [2ex] 
&& 
                - m^2t(6t^2 + 2tu + u^2)
               \Big  \}
            - \frac{2G_5}{s(-4m^2 + s)^2}
               \Big \{-32m^{10} + 16m^8(3s + 4t)  
\nonumber \\ [2ex] 
&& 
                 + m^6(-42st - 32t^2 + 34su) - 
                 4m^4s(4t^2 + 10tu - u^2) + 
                 s(t + u)^2(2t^2 + 2tu + u^2)  
\nonumber \\ [2ex] 
&& 
                + m^2s(2t^3 - 3t^2u - 6tu^2 - u^3)
               \Big  \}
            + \frac{2G_4}{(-4m^2 + s)^2t }
               \Big \{-36m^{10} + 4m^8(13t - 6u)  
\nonumber \\ [2ex] 
&& 
            +     m^2t(2t - 9u)(t + u)^2 + 
                 tu(t + u)^2(2t + 3u) + 
                 m^6(-5t^2 + 66tu + 3u^2)  
\nonumber \\ [2ex] 
&& 
                + m^4(-4t^3 - 49t^2u + 2tu^2 + 3u^3)
              \Big  \}
           + \frac{1}{(-4m^3 + ms)^2t}
           \Big \{  -456m^{12} 
\nonumber \\ [2ex] 
&& 
           + 8m^{10}(55t - 38u) 
             - 9t^2(t - u)u(t + u)^2  
            +   m^8(226t^2 + 648tu + 38u^2)  
\nonumber \\ [2ex] 
&& 
            +   m^4t(9t^3 - 217t^2u - 201tu^2 - 47u^3) - 
               2m^6(90t^3 + 129t^2u - 64tu^2 - 19u^3)  
\nonumber \\ [2ex] 
&& 
            +   2m^2t(18t^4 + 18t^3u + 17t^2u^2 + 
               16tu^3 - u^4)
           \Big  \}
             \Big]  ~, 
\eea

\bea
C_{22}^u &=&         
            \frac{  (4\zeta(2) -9)
            (t - u)}{4 m^2}
              \Big[-20m^6 + 15m^4(t + u) + tu(t + u)  
\nonumber \\ [2ex] 
&& 
             - 4m^2(t^2 + tu + u^2)
              \Big]  ~,  
\\ [4ex] 
C_{33}^u &=&         \frac{2}{4t}
                 \Big[
                 \frac{1}{m^2-t}
                \Big \{-9m^8 - 4m^4t^2 + 3m^6(5t + u)  
\nonumber \\ [2ex] 
&& 
               -  m^2tu(9t + u) + t^2u(2t + 3u)
               \Big  \}
\nonumber \\ [2ex] 
&& 
               - 2(G_7 - G_3)
                \Big \{-6m^6 + 2m^4(8t + u) - m^2t(10t + 7u)  
\nonumber \\ [2ex] 
&& 
              +   t(2t^2 + 2tu + u^2)
               \Big  \}
                 \Big]  ~,
\\ [4ex] 
C_{00}^u &=&       - C_{22}^u   ~.
\eea

where
\bea
G_1 &=& \zeta(2)  \ln \left(\frac{(t-m^2)^2}{m^2s} \right)- 
\ln \left(   {-\frac{t}{m^2}} \right)  ~,   
\nonumber \\ [2ex]
G_2 &=& 2 
             \ln \left(- \frac{t}{m^2} \right)
            \ln \left(\frac{(t-m^2)^2}{m^2s} \right)
            + 4 Li_2 \left( \frac{t}{m^2} \right)
            - \ln^2\left(\frac{-t}{m^2}\right)~ ,
\nonumber \\ [2ex]
G_3 &=& \ln \left(  {-\frac{t}{m^2}}\right)~, \quad
\quad 
G_4 =  \ln \left(  {\frac{s}{m^2}} \right)  ~,~~
\nonumber \\ [2ex]
G_5 &=& \frac{1}{\beta}
            \Big[ \ln^2(\gamma) + 4 Li_2 (-\gamma) +2 \zeta(2)
              \Big] ~,~~
\nonumber \\ [2ex]
G_6 &=&\ln \left( {-\frac{t}{m^2}} \right)  \ln \left(\frac{(t-m^2)^2}{m^2s} \right) + 2 Li_2 \left( \frac{t}{m^2} \right)  ~,
\nonumber \\ [2ex]
G_7 &=& \ln \left(\frac{(t-m^2)^2}{m^2s} \right) ~, \quad \quad
G_8 = \ln ^2 \left(\frac{-t}{m^2} \right) ~,
\nonumber \\ [2ex]
\gamma &=&  \frac{1-\beta}{1+\beta} ~,
\quad  
\beta  =  \sqrt {1- 4m^2/s} ~.
\nonumber \\
\eea

\bea
\!\!\!\! \overline{|M^{V}|^2}_{gg ,int}^{fin} 
&=& 
 {\cal T}_W^2 \frac{2}{C_A} \frac{e^2 \kappa^2}{N^2-1}
   \times \Bigg[
   \Big\{H_1(t) \Big(9m^4 +2t^2 +2tu +u^2 -6m^2(t+u) \Big) 
\nonumber \\ [2ex] 
&&
  +\frac{H_2(t)}{4(t-m^2)^2} 
    \Big(-9m^8 
   +t^2u(2t+u) -2m^2tu(3t+u) +2m^6(5t+3u)
\nonumber \\ [2ex] 
&&
   -m^4(3t^2-2tu+u^2) \Big) 
+ u \leftrightarrow t \Big\}
+\frac{H_3}{4} \Big(18m^4 +3t^2 +4tu +3u^2 -12m^2(t+u) \Big)
\nonumber \\ [2ex] 
&&
+\frac{H_4}{4 (4m^2-s)^2} \Big( -32m^8 +4m^6(t+u) -(t+u)^2(t^2+4tu+u^2) 
\nonumber \\ [2ex] 
&&
+m^4(6t^2+44tu+6u^2) +2m^2(t^3-3t^2u-3tu^2+u^3) \Big)
+\frac{H_5}{s (4m^2-s)^2} 
\nonumber \\ [2ex] 
&&
\Big( 80m^{10} 
-32m^8(t+u) +8m^2tu(t+u)^2 -16m^6(t^2+5tu+u^2) 
\nonumber \\ [2ex] 
&&
           -(t+u)^3(3t^2+4tu+3u^2) +2m^4(5t^3 +31t^2u +31tu^2 +5u^3) \Big)
\nonumber \\ [2ex] 
&&
-\frac{H_6}{4 (t-m^2)^2 (u-m^2)^2} \Big( 18m^{12}-34m^{10}(t+u) -t^2u^2(t^2+4tu+u^2) 
\nonumber \\ [2ex] 
&&
+m^8(25t^2 +36tu+25u^2) 
            +2m^2tu(t^3+6t^2u +6tu^2+u^3) 
\nonumber \\ [2ex] 
&&
-4m^6(2t^3 +t^2u+tu^2+2u^3) +m^4(t^4-8t^3u-20t^2u^2-8tu^3+u^4) \Big)
\nonumber \\ [2ex] 
&&
+\frac{H_7}{4(t-m^2)(u-m^2)(4m^2-s)} \Big( -28m^{10} +8m^6tu +26m^8(t+u) 
\nonumber \\ [2ex] 
&&
+tu(t^3+t^2u+tu^2+u^3) -m^4(5t^3+23t^2u +23tu^2 +5u^3)
\nonumber \\ [2ex] 
&&
            + m^2(t^4 +4t^3u +10t^2u^2+4tu^3+u^4) \Big) \Bigg] 
\eea
where
\bea
\nonumber 
H_1(t) &=&  \frac{1}{8} \Bigg(  2 \ReDs \ln \left(\frac{-t}{m^2} \right) 
                               \ln \left(\frac{(m^2-t)^2}{m^2 s }\right)
\nonumber \\ [2ex] 
&&
                              + 2 \ImDs \pi \ln \left(\frac{(m^2-t)^2}{m^2 s }\right)
                              + 4 \ReDs Li_2{ \left( \frac{t}{m^2}\right)}
\nonumber \\ [2ex] 
&&
                              - \ReDs  \ln \left(\frac{-t}{m^2}\right)^2
                              -2 \ImDs \pi \ln \left(\frac{-t}{m^2}\right)
                     \Bigg),
\nonumber \\ [2ex] 
H_1(u) &=&  \frac{1}{8} \Bigg(  2 \ReDs \ln \left(\frac{-u}{m^2} \right) 
                               \ln \left(\frac{(m^2-u)^2}{m^2 s }\right)
\nonumber \\ [2ex] 
&&
                                + 2 \ImDs \pi \ln \left(\frac{(m^2-u)^2}{m^2 s }\right)
                                + 4 \ReDs Li_2{ \left( \frac{u}{m^2}\right)}
\nonumber \\ [2ex] 
&&
                                - \ReDs  \ln \left(\frac{-u}{m^2}\right)^2
                                -2 \ImDs \pi \ln \left(\frac{-u}{m^2}\right)
                     \Bigg),
\nonumber \\ [2ex]
H_2(t) &=&   \ReDs \ln  \left( \frac{-t}{m^2} \right), 
\quad \quad \quad \quad
H_2(u) =   \ReDs \ln \left( \frac{-u}{m^2} \right), 
\nonumber \\ [2ex]
H_3 &=&  \zeta(2) \ReDs,
\quad \quad \quad \quad 
H_4 =  \ReDs ln \left( \frac{s}{m^2} \right),
\nonumber \\ [2ex]
H_5 &=&  \frac{1}{8} \Big(  \frac{\ReDs}{\beta} \ln^2 {\left(\gamma\right)} +
                            \frac{4}{\beta} \ReDs Li_2 {\left( -\gamma \right)} +
                            \frac{2 \ReDs}{\beta} \zeta(2)
                     \Big),
\nonumber \\ [2ex]
H_6 &=&  \ImDs \pi,
\quad \quad \quad \quad
H_7 =  \ReDs .
\nonumber \\
\eea

%
%
%
%
%
%
%
%
%
%


\end{document}